\documentclass[review]{elsarticle}
\usepackage{graphics}
\usepackage{graphicx}
\usepackage{float}
\usepackage{bm}
\usepackage[nomarkers,notablist,nofiglist]{endfloat}
\usepackage{lineno,hyperref}
\modulolinenumbers[5]
\usepackage{siunitx}         
\graphicspath{ {./Figures/} } 
\usepackage{soul}
\usepackage{amsmath}
\usepackage{geometry} 
\usepackage{color} 
\usepackage{outlines}
\usepackage{textcomp}
\journal{arXiv}

\begin{document}

\begin{frontmatter}

	\title{Combining Synchrotron X-ray Diffraction, Mechanistic Modeling, and Machine Learning for \emph{In Situ} Subsurface Temperature Quantification during Laser Melting}

	\author[1]{Rachel E. Lim}

	\author[1]{Tuhin Mukherjee}
	\author[2]{Chihpin Chuang}
	\author[3]{Thien Q. Phan}
	\author[1]{Tarasankar DebRoy}
	\author[1]{Darren C. Pagan\corref{mycorrespondingauthor}}
	\ead{dcp5303@psu.edu}

	\cortext[mycorrespondingauthor]{Corresponding author}

	\address[1]{Pennsylvania State University, University Park, PA 16802, USA}
 	\address[2]{Argonne National Laboratory, Lemont, IL 60439, USA}
	\address[3]{Lawrence Livermore National Laboratory, Livermore, CA 14850, USA}

	\begin{abstract}
		Laser melting, such as that encountered during additive manufacturing (AM), produces extreme gradients of temperature in both space and time, which in turn influence microstructural development in the material. Qualification and model validation of the process itself and resulting material produced necessitates the ability to characterize these temperature fields. However, well-established means to directly probe material temperature below the surface of an alloy while it is being processed are limited. To address this gap in characterization capabilities, we present a novel means to extract subsurface temperature distribution metrics, with uncertainty, from \emph{in situ} synchrotron X-ray diffraction measurements to provide quantitative temperature evolution during laser melting. Temperature distribution metrics are determined using Gaussian Process Regression supervised machine learning surrogate models trained with a combination of mechanistic modeling (heat transfer and fluid flow) and X-ray diffraction simulation. Trained surrogate model uncertainties are found to range from 5\% to 15\% depending on the metric and current temperature. The surrogate models are then applied to experimental data to extract temperature metrics from an Inconel 625 nickel superalloy wall specimen during laser melting. Maximum temperatures of the solid phase in the diffraction volume through melting and cooling are found to reach the solidus temperature as expected, with mean and minimum temperatures found to be several hundred degrees less. The extracted temperature metrics near melting are determined to be more accurate due to the lower relative levels of mechanical elastic strains. However, uncertainties for temperature metrics during cooling are increased due to the effects of thermomechanical stress.
	\end{abstract}


\end{frontmatter}





\section{Introduction}

Primary factors for controlling the microstructure, porosity, and residual stress state during additive manufacture (AM) of engineering alloys include  heat input, resulting temperature, and temperature gradients through the component. Heat-treating of components using the heating sources themselves is also of increasing importance. In response, significant efforts have been undertaken to characterize temperature profiles during AM builds in order to guide the build design process. These efforts include both thermal and optical imaging of the specimen surfaces during a build to estimate both surface temperature and melt-pool shape \cite{moylan2014infrared,fox2017measurement,fisher2018toward,montazeri2019heterogeneous,dunbar2018assessment,forien2020detecting,ashby2022thermal}, along with predicting defect formation. While valuable, these characterization efforts only provide information about the temperature profile at the sample surface, precluding understanding critical subsurface thermal profiles during initial melting and subsequent reheating events as layers are added above a volume of material of interest. Rather, different measurement modalities are needed to probe an alloy's evolving microstructure during repeated thermal cycling encountered during the build process.

To address these challenges, new synchrotron X-ray imaging and diffraction capabilities have been developed that can characterize structure at rapid time scales ($\ll s$). These measurements probe subsurface thermomechanical state (thermal and mechanical lattice strain) and microstructure evolution in conditions mimicking a wide range of additive manufacturing processes \cite{Kenel2016,Calta2018,Cunningham2019,Hocine2020,oh2021high,oh2021microscale,thampy2020subsurface}. During these experiments, average temperatures of the crystalline phases within a diffraction volume are estimated from the shifts in diffraction peak centroid positions due to a convolution of microstructure evolution, thermal, and mechanical strains. The effects of stress (elastic strains) and microstructure evolution (such as changes in local composition or precipitation) are often neglected.  With knowledge of the coefficient of thermal expansion (CTE) of the material within the applicable range of temperatures, and assuming equilibrium CTEs are valid during rapid cooling, thermal strains are mapped directly to temperatures. While valuable, accuracy of temperatures from this peak centroid analysis can be compromised due to the inherent spatial gradients of temperature, mechanical loading, and chemistry during the build process. However, while a complication for data analysis, information regarding the spatial gradients of temperature  (along with mechanical strains and chemistry variation) within a diffraction volume is encoded into each diffraction peak. Unfortunately, the single projection of X-rays through the diffraction volume during these experiments prevents the direct inversion or reconstruction of the temperature field in a tomographic fashion.


Here we propose a novel path forward to extracting this temperature data in which a mechanistic heat transfer and fluid flow model and X-ray simulations provide a framework for interpreting and analyzing complex experimental temperature distributions. The simulations are used to create a collection of reference diffraction patterns (a `dictionary') representing different thermal states. A strategy is then adapted from \cite{Bamney2020} to use Gaussian Process Regression (GPR) to learn a mapping between spatial temperature distribution metrics within a diffraction volume (generated from mechanistic modeling) and diffraction peak shapes. The GPR approach taken here is a transfer learning process, in which GPR is applied to simulated training datasets and then the `learned' relationships between spatial temperature distribution metrics within a diffraction volume and diffraction peak shapes are then transferred to experimental datasets of interest. The simulation includes both heat transfer and fluid flow modeling coupled with X-ray diffraction modeling to generate synthetic diffraction datasets. The accuracy of GPR process as used here depends on how well it can predict the temperature field \emph{outputs} (descriptors of the temperature fields present) given particular diffraction \emph{inputs} (diffraction data). Note that since training is performed with synthetic datasets, the temperature fields used to create the diffraction datasets are known. As such, the accuracy of the diffraction modeling is more critical than the heat transfer and fluid flow modeling (although if the heat transfer and fluid flow modeling is accurate, uncertainties will be decreased). Fortunately, the physics of X-ray diffraction are well understood. When applying the trained GPR models to experimental data, uncertainties reflect differences between the training and testing data.


\section{Material and Experiment Description}
\label{sec:expdesc}

\begin{figure}[h!]
    \centering \includegraphics[width=0.7\textwidth]{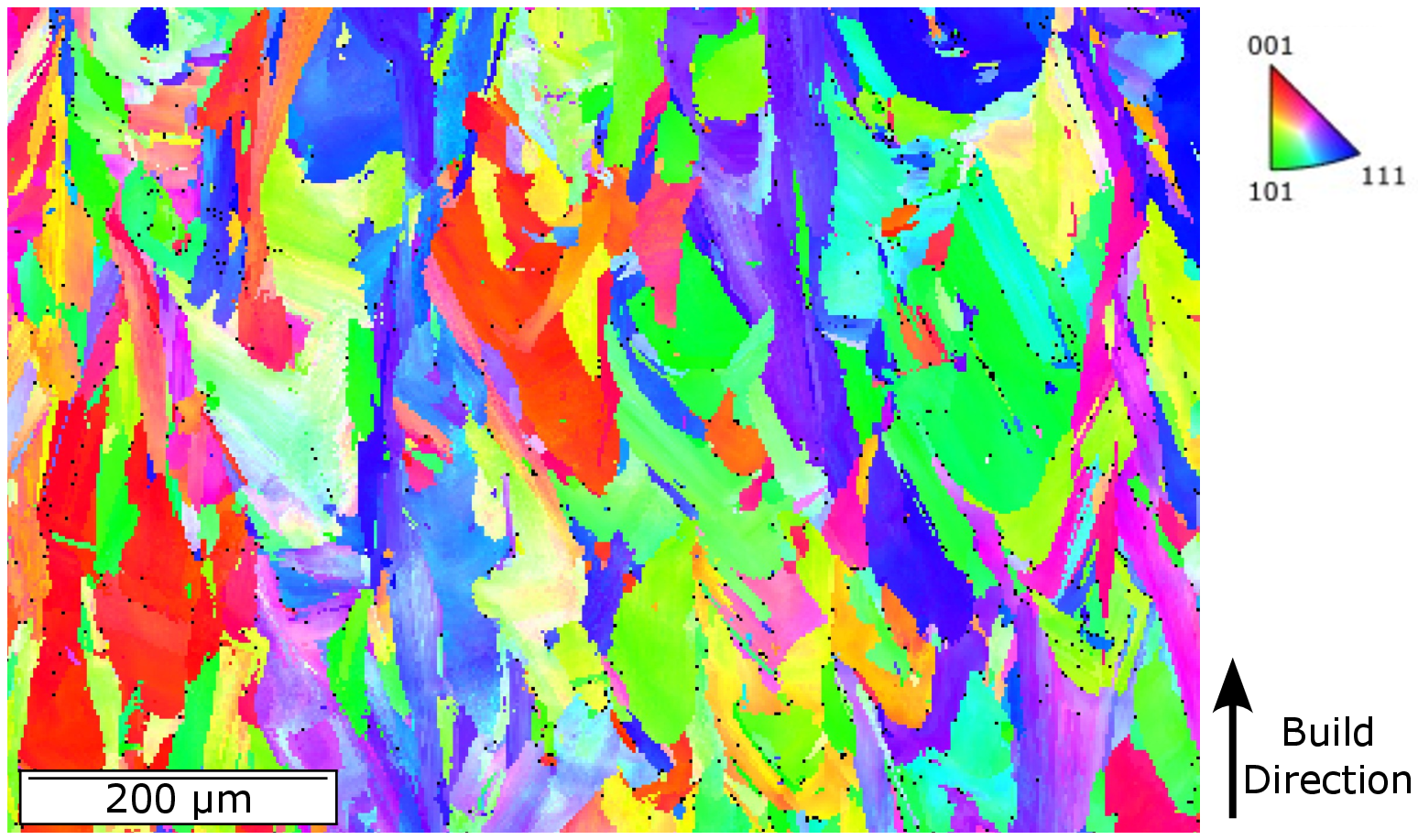}
    \caption{A representative orientation map measured using electron backscatter diffraction from the thin wall specimen tested in this work. Crystal directions are colored with respect to the build direction in the provided inverse pole figure color map.}
    \label{fig:ebsd}
\end{figure}

The specimens in both experiment and simulations used in this work were AM Inconel 625 (IN625) nickel superalloy made using laser powder bed fusion (LPBF) at the National Institute of Standards and Technology (NIST), designation PBF-LB-IN625. The experimental specimen was a thin wall 3 mm in height along the build direction ($\bm{z^S}$) by 0.53 mm in thickness~($\bm{y^S}$) and 20 mm in length ($\bm{x^S}$). The experimental specimen was extracted using electro discharge machining from a block (50 mm $\times$ 15 mm $\times$ 6 mm) built in an EOS M290 machine using manufacturer recommended build parameters \cite{son2020creep}. The IN625 powder used for the build was attained from the machine manufacturer EOS. The build layer thickness was 40 $\mu$m, with a 110 $\mu$m hatch spacing. The laser power and speed were 285 W and 960 mm/s respectively. The interlayer rotation during the build of the larger block from which the thin wall specimen was extracted was 67.5$^\circ$. After build and prior to thin wall specimen extraction, the block was stress-relief heat treated at 800~$^\circ$C for 2 hr. The IN625 used in this work was built using the same machine, powders, and build parameters of material utilized for the NIST AM Bench 2018 challenge \cite{levine2020outcomes}. An orientation map measured using electron backscatter diffraction from the thin wall specimen prior to laser remelting  is shown in Fig. \ref{fig:ebsd}. Crystal directions are colored with respect to the build direction using an inverse pole figure color map. The microstructure primarily consists of large grains with dimensions on the order of 100 $\mu$m, interspersed with smaller grains on the order of 10 $\mu$m.

\begin{figure}[h]
      \centering \includegraphics[width=0.65\textwidth]{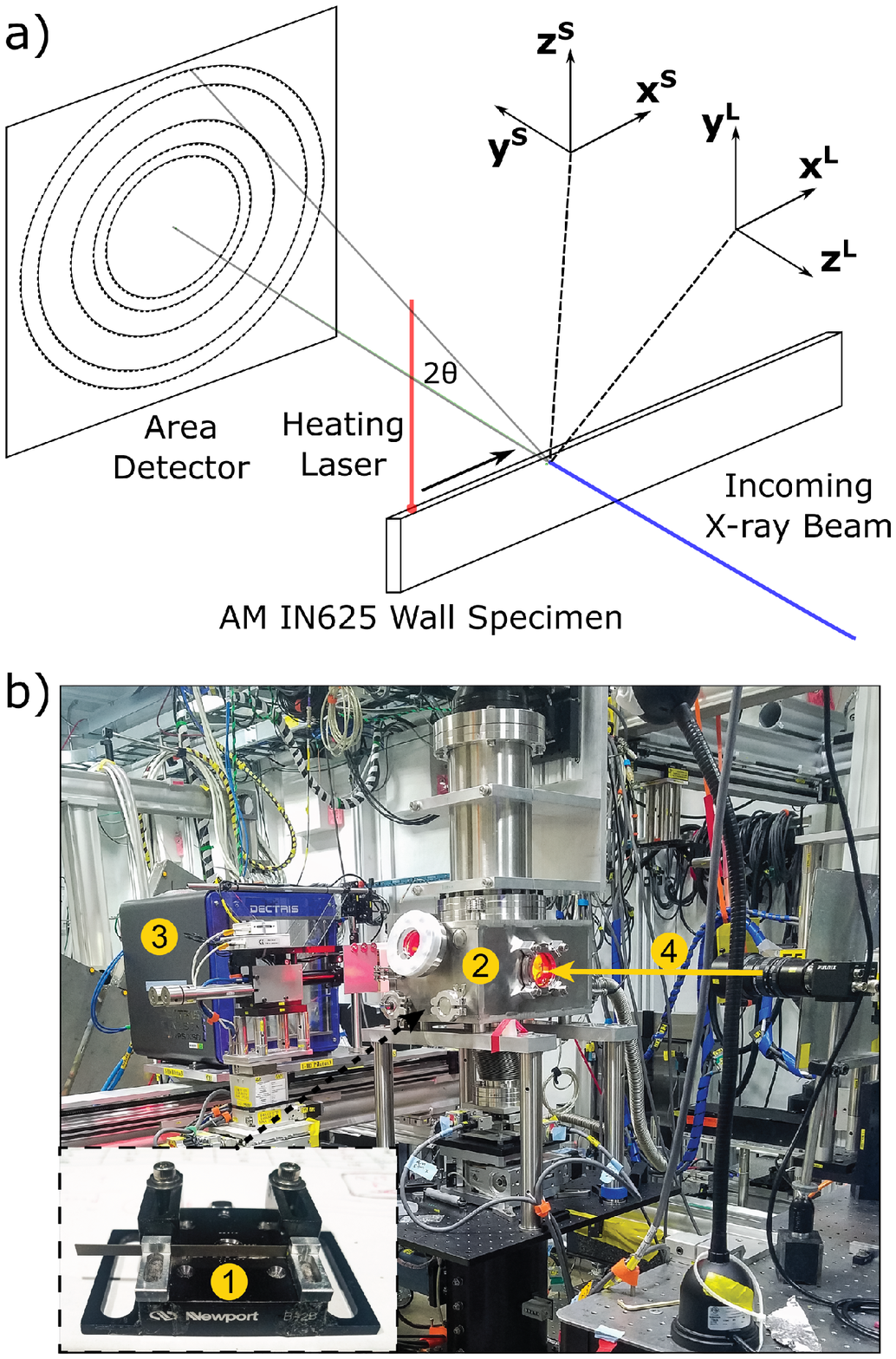}
      \caption{a) Schematic of the experimental setup with `S' superscripts for the sample reference frame associated with the heat transfer and fluid flow modeling and `L' superscripts for the laboratory reference frame associated with the laser heating and with the X-ray diffraction simulations. b) Photo of the experimental measurement set-up used to collect X-ray diffraction data for developing the temperature extraction framework. Marked are (1) the sample holder and sample, (2) the laser test system, (3) X-ray area detector, and (4) incoming beam direction.}
      \label{fig:exp_setup}
\end{figure}

The \emph{in situ} diffraction measurements during laser melting were performed at beamline 1-ID at the Advanced Photon Source (APS). Fig. \ref{fig:exp_setup}a shows a schematic of the experimental geometry for the measurements including AM IN625 wall specimen, heating laser, and incoming X-ray beam, and the orientations of the sample (`S') and laboratory (`L') coordinate system. In the laboratory coordinate system, the incoming X-ray beam travels in the $-\bm{z^L}$ direction while the heating laser was nominally aligned along $\bm{y^L}$. During X-ray measurements, the specimen remained fixed in the laboratory coordinate system as the laser traveled in the $\bm{x^L}$ / $\bm{x^S}$ direction. The angle between incoming and diffracted X-rays $2\theta$ is labeled and is related to the spacing of diffracting sets of lattice planes. The incoming X-ray beam was 61.332 keV and was focused vertically by a set of Si sawtooth lenses to dimensions of 50 $\mu$m $\times$ 30 $\mu$m along $\bm{x^L}$ and $\bm{y^L}$ respectively. X-ray diffraction images were measured by a Pilatus3 X CdTe 2M detector sitting 752 mm downstream of the sample. The detector has 1475$\times$1679 pixels and a pixel size of 172$\times$172 ${\mu m^2}$. Diffraction images were collected with an exposure time of 1 ms and a frequency of 250 Hz throughout the experiment.

\begin{figure}[h!]
    \centering \includegraphics[width=0.8\textwidth]{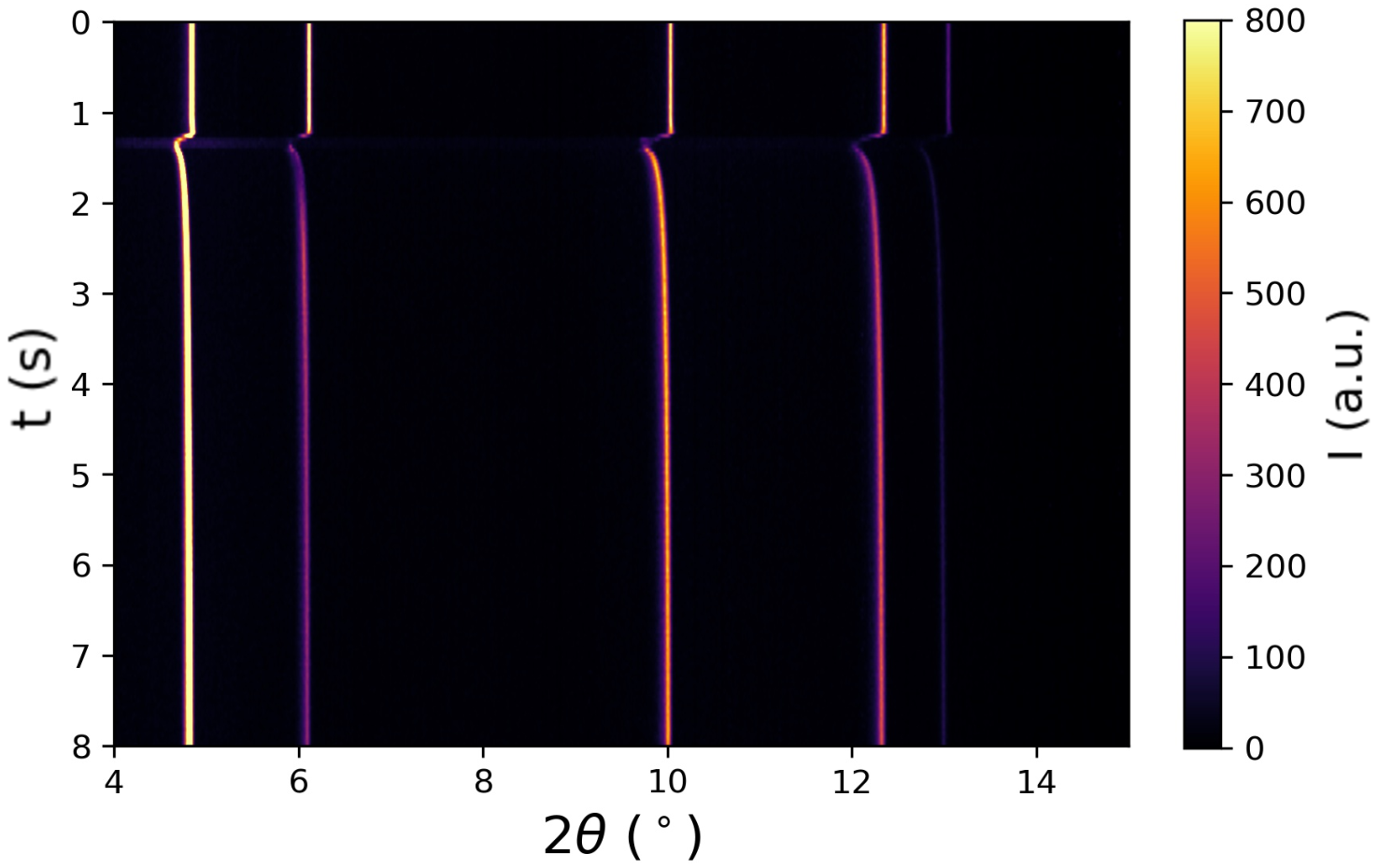}
    \caption{Evolution of azimuthally integrated diffracted intensity $I$ with time $t$ during laser melting of the IN625 wall specimen. Each row corresponds to the diffraction line profile (intensity vs 2$\theta$) for a given time step with the color signifying the magnitude of diffracted intensity.}
    \label{fig:heatmap}
\end{figure}

Laser melting was performed using an existing \emph{in situ} LPBF simulator at the APS. A picture of the simulator in Sector 1-ID is shown in Fig. \ref{fig:exp_setup}b and more complete system details can be found in \cite{Zhao2017}. The LBPF simulator uses a ytterbium fiber laser (IPG YLR-500-AC, USA) and an $intelliSCAN_{de}$\textsuperscript{\textregistered} 30 for laser motion. Prior to laser melting, the environment chamber was purged and re-filled with high-purity argon gas. During testing, the specimen was placed 2.9 mm away from the laser focal plane to create a spot diameter of 100 $\mu$m. During X-ray diffraction measurements, the laser was rastered over the wall specimen along $\bm{x^L}$ with a laser power of approximately 120 W and speed of 0.05 m/s. We note that this linear power density (2,400 J/m) is relatively high in comparison to standard LPBF parameters for IN625. These parameters were chosen to ensure a relatively large melt-pool and extended temperature gradient through the thickness of the specimen. Figure \ref{fig:heatmap} shows the diffracted intensity integrated azimuthally around the detector (along the diffraction rings) vs time. In the figure, we can see the shifting of the diffracted intensity to lower $2\theta$ at 125 ms. This shift to lower $2\theta$ is due to an increase in lattice plane spacing as the diffraction volume is rapidly heated due to the laser passing over the diffraction volume.

\section{Methods} 

\begin{figure}[h!]
    \centering \includegraphics[width=1.0\textwidth]{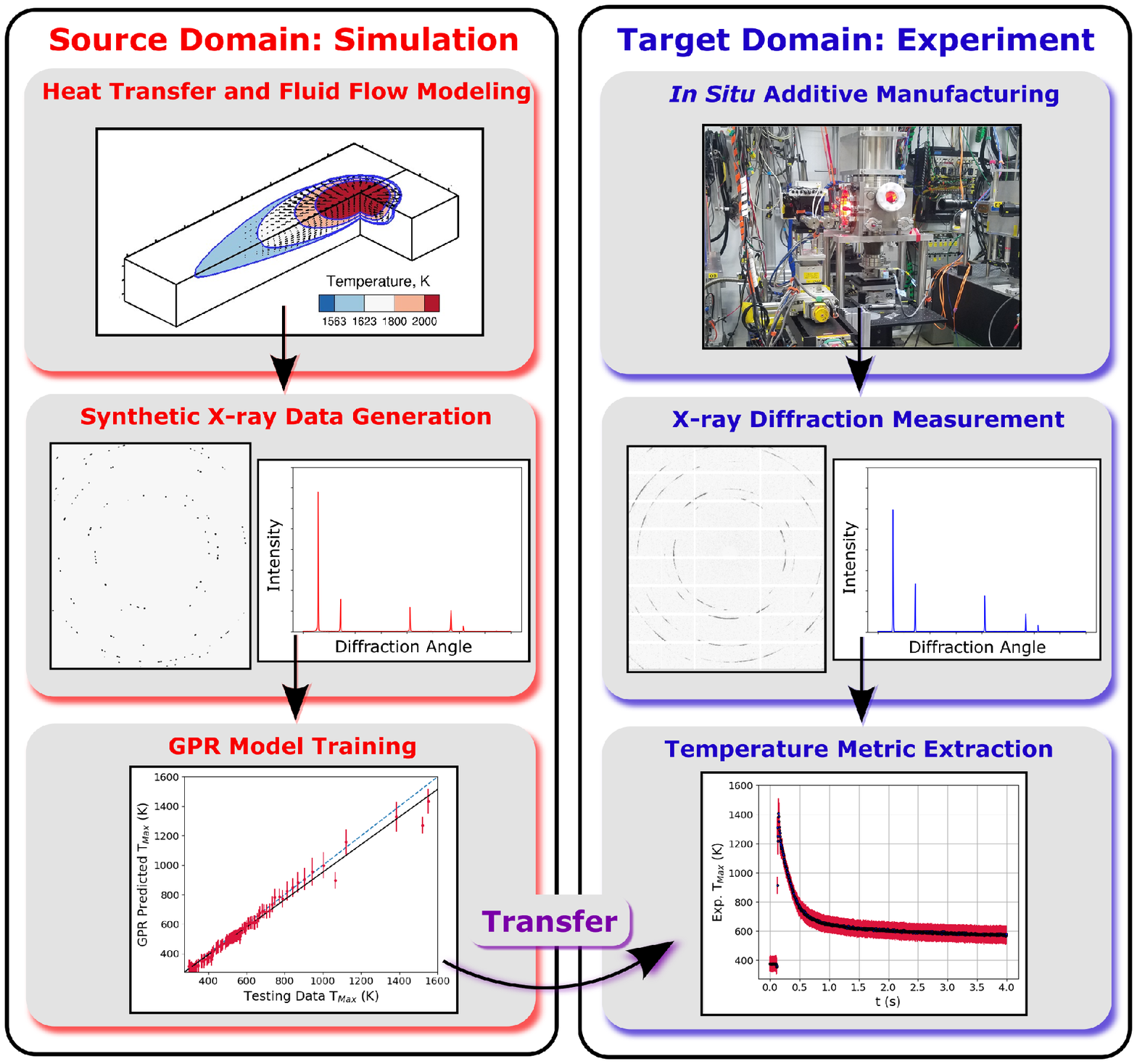}
    \caption{Flow chart showing the steps included in this work to extract temperature metrics from X-ray diffraction data. Within the Source Domain, heat transfer and fluid flow modeling is used to inform X-ray diffraction simulations which are then used to train GPR models. The trained GPR models are transferred to the Target Domain and used to predict temperature metrics from the X-ray experimental data.}
    \label{fig:overview}
\end{figure}

In this section, an overview of the various methods employed to build the dictionary of reference diffraction patterns linked to underlying thermal distributions is given. A description of the GPR model, training, and hyperparameter selection used to learn the mapping between underlying temperature distribution metrics and diffraction line profiles is also provided. As previously described in the Introduction, as opposed to using X-ray diffraction data to validate a simulation, we are using combinations of heat transfer and fluid flow modeling, X-ray diffraction simulation, and machine learning to interpret and extract information from the experimental data. A schematic of the various components of the effort are shown in Fig. \ref{fig:overview} and the components of the modeling and training effort are described in more detail in the following subsections. As a short summary, synthetic X-ray data are generated using thermal fields output from the mechanistic heat transfer and fluid flow model. We adopt an approach outlined in Fig. \ref{fig:overview} in which pairs of synthetic X-ray line profiles (input) and underlying temperature metrics (output) are used to train various GPR surrogate models. Then, the surrogate models, trained using synthetic data (the Source Domain), are \emph{transferred} \cite{weiss2016survey} to analyze experimental data (the Target Domain) to determine temperature metrics through, essentially, comparisons with the previously generated synthetic data.


\subsection{Heat Transfer and Fluid Flow Modeling}

\label{sec:thermal}

A heat transfer and fluid flow model is utilized to calculate the 3D transient temperature and velocity fields during the laser melting of IN625 specimens. The model is discussed in detail in Mukherjee et al. \cite{Mukherjee2018,Mukherjee2018a} and only the important features are described here. The model was developed using an in-house Fortran code and compiled with an Intel Compiler. The model calculates the melt-pool size, temperature fields, and velocity fields during the LPBF process taking the laser parameters, alloy, and environmental gas properties as inputs. The model considers temperature-dependent thermophysical properties for both powder and fully dense material. The thermophysical properties of IN625 required in the model are calculated using the commercial package JMatPro, while the uncertain parameters such as absorptivity and the power distribution of the laser beam can be adjusted to match the experimental data on thermal cycles and deposit geometry. The material parameters used for the heat transfer and fluid flow modeling are provided in Table \ref{tab:in625_SimProps}. The model iteratively solves the equations of conservation of energy, mass, and momentum in a 3D computational domain consisting of the substrate, power bed, deposited layers and hatches, and the shielding gas. The equations are discretized in the computational domain using a finite difference scheme and a traveling grid approach is used to increase computational efficiency.  The model provides accurate results on melt-pool geometry and temperature fields by considering the effects of the convective flow of molten metals. We mention that while laser melting and resulting temperature distributions within solid specimens are modeled in this work, the same model is capable of modeling temperature distributions within loose powder layers \cite{Mukherjee2018a}.

\begin{table}[h]
    \begin{center}
        \centering
        \caption{Properties of Inconel 625 used in the heat transfer and fluid modeling. These properties represent the thermo-
physical behavior of the alloy and affect the thermal cycles. Here, thermal conductivity and
specific heat are taken as temperature dependent and the temperature in K is represented by $T$.
The properties were calculated using the commercial software, JMatPro.}
        \label{tab:in625_SimProps}
        \begin{tabular}{ m{8.2cm} | m{5cm} } 
            \textbf{Physical Property} & \textbf{Value}  \\
            \hline
            Density (\si{kg/m^{3}}) & 8440 \\
            Solidus temperature (\si{K}) & 1563 \\
            Liquidus temperature (\si{K}) & 1623 \\
            Specific heat (\si{J/kg/K}) & $360.4+0.26~T - 4\times 10^{-5}~T^2$ \\
            Thermal conductivity (\si{W/m/K}) & $0.56 + 2.9 × \times 10^{-2}~T - 7 \times 10^{-6}~T^2$\\
            Latent heat of fusion (\si{J/kg}) & 209.2 $\times 10^3$ \\
            Viscosity (\si{kg/m/s}) & 5.3 $\times 10^{-3}$ \\
            Temperature coefficient of surface tension (\si{N/m/K}) & -0.37 $\times 10^{-3}$ \\
            Surface Tension (\si{N/m}) & 1.82 \\
            Absorptivity factor & 0.3 \\
            Emissivity factor & 0.4 \\
        \hline
    \end{tabular}
    \end{center}
\end{table}

Using this model, a series of single laser trace simulations were performed around conditions similar to the previously performed \emph{in situ} synchrotron experiments (\S \ref{sec:exp_data}).  For all simulations, a unidirectional scan along $\bm{x^S}$  of the laser beam is used, with the laser beam direction being $-\bm{z^S}$. Positive $\bm{y^S}$ is perpendicular to the laser scanning direction and represents the direction along the width of the wall specimen.  The laser power and velocity were varied around the nominal experimental parameters (120 W laser power and 50 mm/s laser speed). In total, nine laser melting simulations were performed according to the simulation test matrix in Table \ref{tab:test_mat}.  Each simulation captured the laser traveling over a 200 ms interval with 200 µs time steps while temperature fields were output every 2 ms. The heat transfer and fluid flow simulations were performed on the ROAR supercomputer at Penn State using 40 cores (2.13 GHz) for each simulation, with each thermal simulation taking one hr to complete.

\begin{table}
\centering
\begin{tabular}{c|c|c}
P = 100 W & P = 100 W & P = 100 W \\ 
v = 0.04 m/s & v = 0.05 m/s & v = 0.06 m/s \\ \hline
P = 120 W & P = 120 W & P = 120 W \\  
v = 0.04 m/s & v = 0.05 m/s & v = 0.06 m/s \\ \hline
P = 140 W & P = 140 W & P = 140 W \\  
v = 0.04 m/s & v = 0.05 m/s & v = 0.06 m/s \\ 
\end{tabular}
    	 \caption{Test matrix showing the nine sets of power-velocity parameters which were used in the thermal simulation.}
    	 \label{tab:test_mat}
\end{table}

\subsection{Synthetic X-ray Data Generation}

Each of the nine laser trace simulations was used to generate a series of time-dependent area detector X-ray diffraction patterns using the framework developed in Pagan et al. \cite{Pagan2020}. This simulator uses diffraction calculation and projection algorithms contained in the Python-based HEXRD software package \cite{Bernier2011}. In the diffraction simulation framework, X-ray diffraction is simulated within a polychromatic (Laue) diffraction framework that employs a finite-energy bandwidth to capture realistic beam conditions. This method provides benefits over angular based diffraction solution methods in conditions where the specimen is not rotating, such as the single laser trace experiments described above. Here the average X-ray energy and bandwidth ($|\Delta E|/E$) were chosen to match the experiment, 61.332 keV and of 5$\times 10^{-4}$ respectively.

In this framework, diffraction events are simulated from discretized volumes in space, with the spatial positions of each volume being incorporated into diffracted ray tracing calculations. For clarity, we will refer to these discretized volumes as scattering volumes, while the total volume illuminated by the X-ray beam is the diffraction volume. Inside each scattering volume, individual grains (lattice orientations) are inserted from which diffraction events are calculated. In these diffraction simulations, the grains have no morphological features and their orientations are randomly generated. The thermomechanical state of each scattering volume, and the accompanying changes in lattice state, can vary spatially. Here, temperature fields produce spatially varying thermal strain within the diffraction volume. Matching the experiment, the simulated diffraction volume is 50 $\times$ 30 $\times$ 530 $\mu$m$^3$, while each scattering volume was 20 $\times$ 20 $\times$ 20 $\mu$m$^3$. Each scattering volume contained two randomly oriented grains simulating grains with approximately 25 $\mu$m equivalent diameter. Each grain contains 1$^\circ$ of lattice misorientation to provide some amount of diffraction peak broadening.

\begin{figure}[h]
  \centering \includegraphics[width=0.6\textwidth]{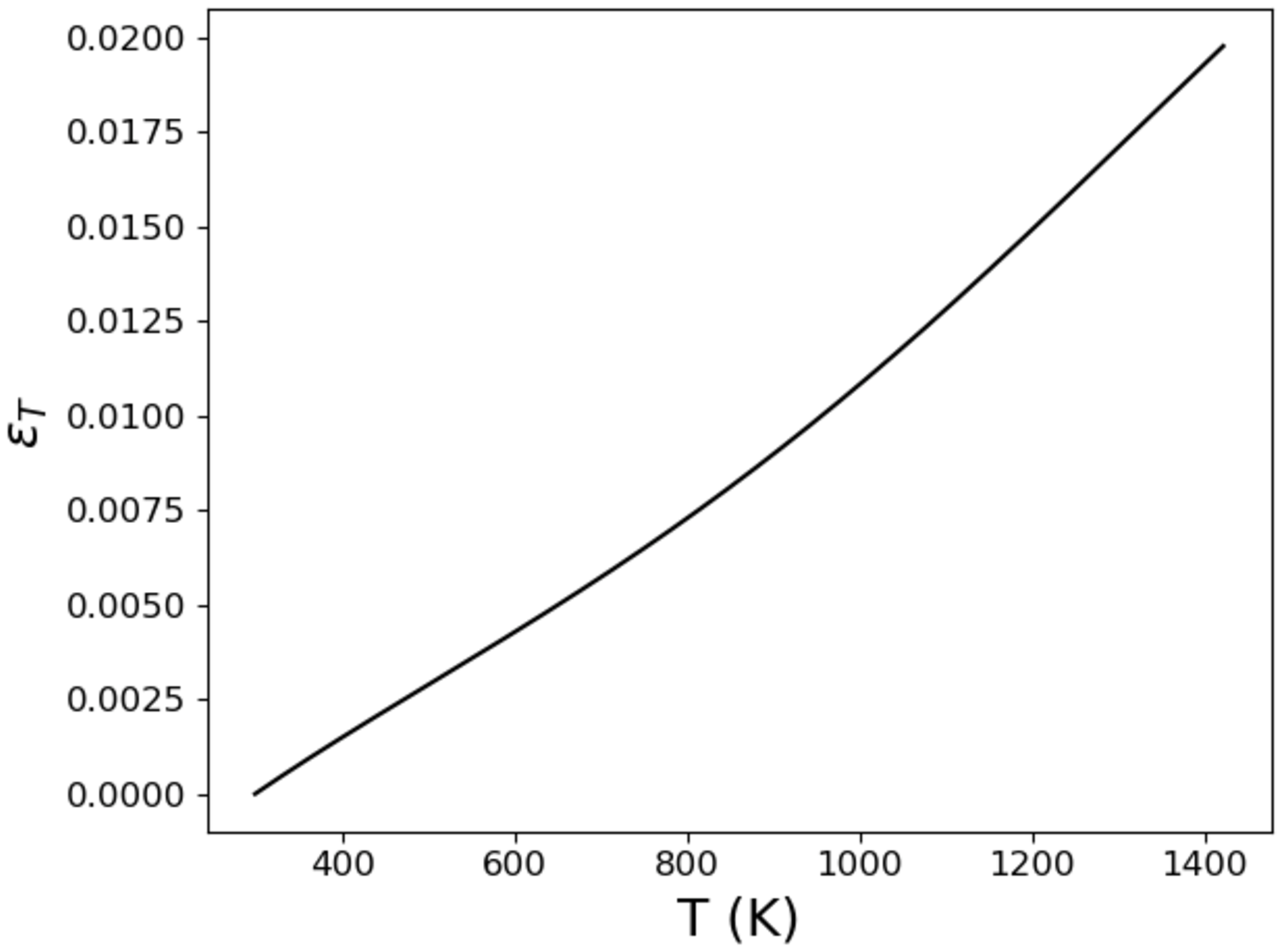}
  \caption{Measured thermal strain $\varepsilon_t$ versus temperature $T$ of IN625 used for the diffraction simulations. Measurements were made using dilatometry on the same material as tested in the synchrotron experiment.}
  \label{fig:thermal}
\end{figure}

To model the thermal strain's effects on the measured diffraction data, the lattice structure of an embedded grain is altered by stretching the reciprocal lattice vectors, $\bm{g}$, of each grain isotropically (valid for a cubic crystal):

\begin{equation}
\bm{g}=(1-\varepsilon_T) \underline{\bm{I}} \cdot \bm{g_0}
\end{equation}
where $\bm{g_0}$ is the unstrained reciprocal lattice vector in a crystal, $\underline{\bm{I}}$ is the second order identity tensor, and the thermal strain $\varepsilon_T$ is given by

\begin{equation}
\varepsilon_T=\int_{T_0}^{T} \alpha(T) \,dT.
\end{equation}
The temperature-dependent coefficient of thermal expansion $\alpha(T)$ used in this work was measured independently using bulk dilatometry measurements at the Penn State Center for Innovative Sintered Products. Bear in mind, these measurements describe equilibrium thermal expansion behavior which may not exactly capture thermal expansion during rapid heating and cooling. The sample used for dilatometry measurements was extracted via EDM from an identically built AM bulk sample used for the \emph{in situ} diffraction experiments.  Thermal strain $\varepsilon_T$ as a function of temperature determined from these measurements and used for diffraction simulations are provided in Fig. \ref{fig:thermal}. The reference lattice parameter from which $\bm{g_0}$ were generated was 3.5981 $\AA$. As only crystalline material will diffract, if the temperature in a scattering volume exceeds the solidus temperature (1563 K \cite{pawel1985survey}), no diffraction events are recorded. For the high-energy transmission geometry used in the experiment, all diffracting X-rays have nearly the same path length regardless of whether they diffract from the upstream or downstream side of the specimen. As such, absorption is not considered since it will only scale the total integrated intensity of the diffraction peak and not change the peak shape.

\begin{figure}[h]
      \centering \includegraphics[width=1.0\textwidth]{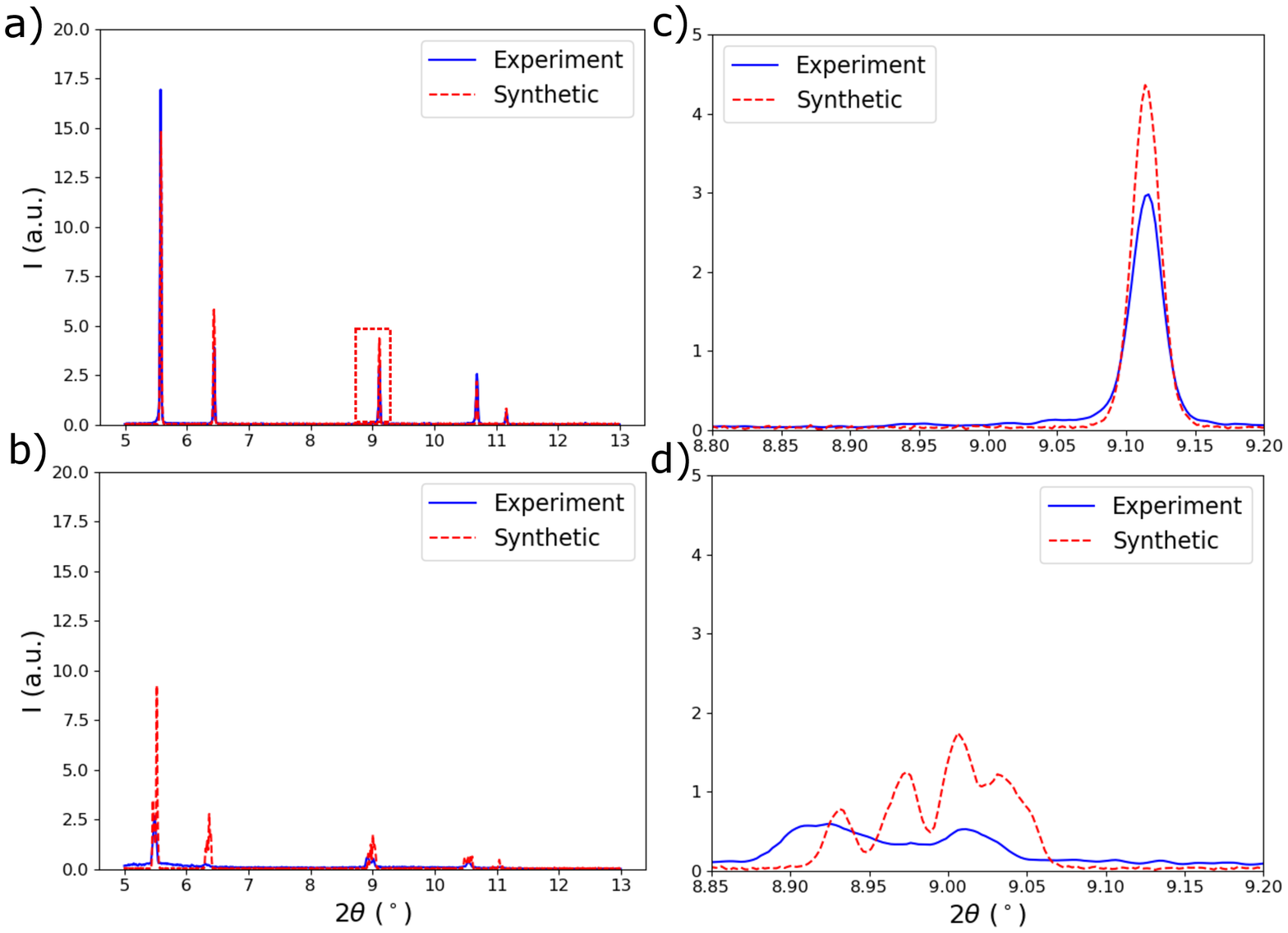}
      \caption{Comparison between experimental (blue) and synthetic (dashed red) diffraction line profiles (intensity $I$ versus Bragg Angle $2\theta$) from the AM IN625 wall specimens in representative a) unheated and b) heated conditions. Enlarged views of the (220) diffraction peak in the c) unheated and d) heated condition. The (220) peak provides an example of relatively extreme peak splitting due to the temperature gradient presented. }
      \label{fig:ex_comp}
\end{figure}

The laser-trace simulations described in \S \ref{sec:thermal} are used as input for generating synthetic diffraction images. As the thermal simulations use a traveling grid formulation for calculation, thermal fields at each time step were mapped to a regular grid of scattering volumes with 20 $\mu$m spacing in all three directions. Each thermal simulation was used to generate three sets of synthetic diffraction data capturing the evolution of different temperature gradients by placing the X-ray beam at different portions of the sample. For each thermal simulation time series, the center of the X-ray beam was placed 20, 40, and 60 $\mu$m below the top of the specimen and separate 2D X-ray diffraction image sets were generated. In total, 27 sets of X-ray simulations and 2700 diffraction images total were generated for surrogate model training. Once the 2D diffraction patterns were simulated for the entire time series, each pattern was integrated azimuthally around the diffraction rings to create 1D diffraction line profiles (intensity vs. 2$\theta$). The integrated diffraction line profile data covers 2$\theta$ angles ranging from $5^\circ$ to $13^\circ$ and encompasses the first six sets of lattice planes (111), (200), (220), (311), (222), (400). After integration, background noise corresponding to scattering in the experimental station was added to the synthetic data. A comparison of example experimental (blue) and synthetic (dashed red) diffraction line profiles in the unheated conditions is given in Fig. \ref{fig:ex_comp}a and soon after laser heating in Fig. \ref{fig:ex_comp}b. Differences in relative peak heights are likely due to local texture in the thin wall sample. Figure \ref{fig:ex_comp}c and Fig. \ref{fig:ex_comp}d show enlarged views of the (220) diffraction peak in the unheated and heated conditions respectively. Relatively extreme peak broadening and splitting due the temperature gradient present can be observed in both the experimental and synthetic diffraction images. However, we note here that the goal of the diffraction simulations is not to exactly match the experimental diffraction line profiles, but to provide a reference dictionary which a trained GPR surrogate model can utilize to predict a temperature metric based on \emph{similar} features found in the data of interest. Each diffraction simulation of a heating time series at a single beam position took approximately 4 hr to complete, with diffraction calculations from each scattering volume parallelized over 40 cores.

\subsection{Gaussian Process Regression Surrogate Model Description}

\begin{figure}[h]
  \centering \includegraphics[width=0.6\textwidth]{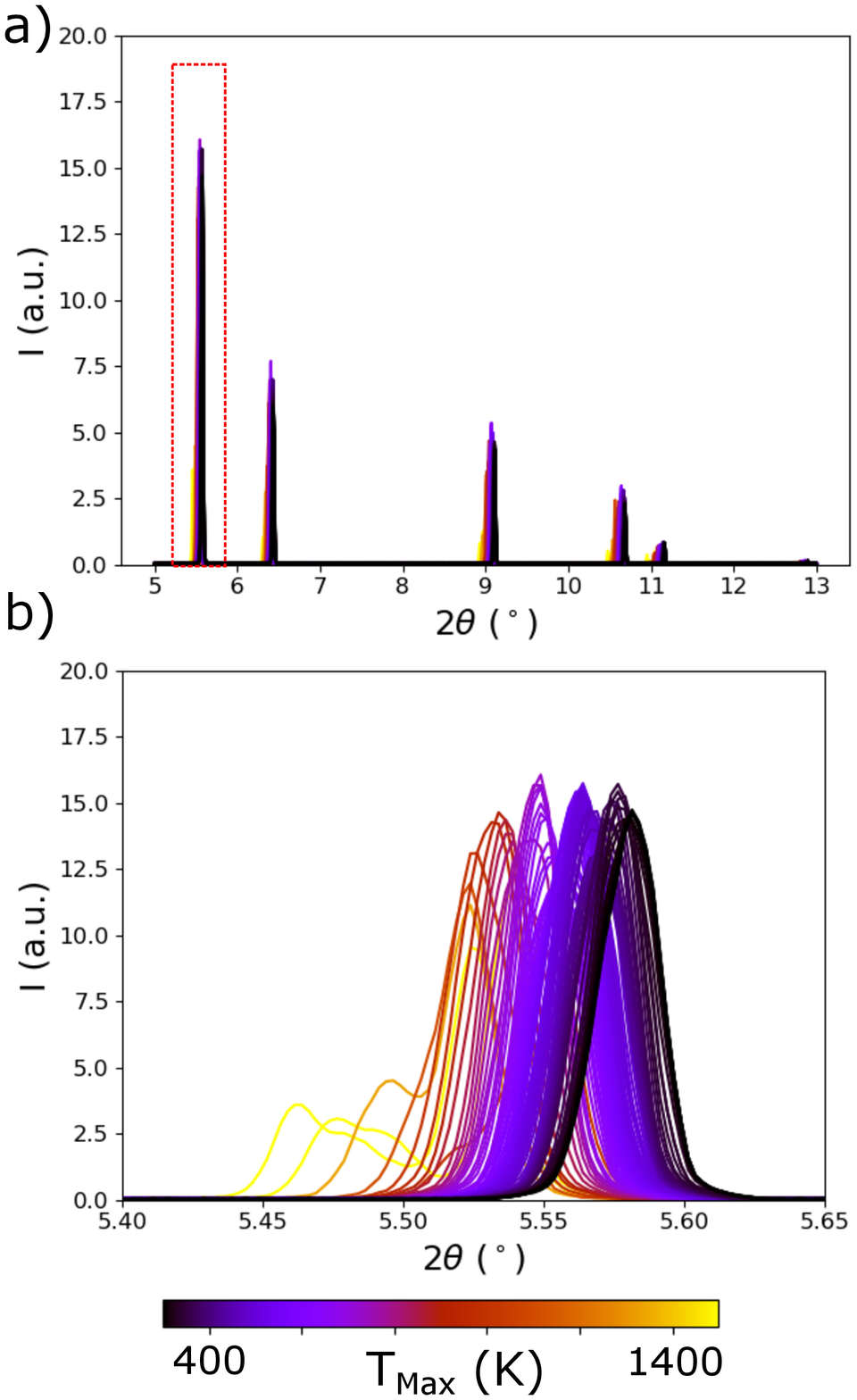}
  \caption{Simulated a) diffraction line profiles (intensity $I$ versus Bragg Angle $2\theta$) and b) (111) diffraction peaks of IN625 colored by the maximum temperature ($T_{\mathrm{Max}}$) in the diffraction volume which was used to generate the pattern.}
  \label{fig:mapping}
\end{figure}

Here we utilize the supervised machine learning technique, Gaussian Process Regression \cite{Williams2006} implemented via scikit-learn \cite{pedregosa2011scikit}, to learn a mapping between diffraction line profiles and various temperature metrics in the diffraction volume. Figure \ref{fig:mapping} shows example synthetic diffraction patterns generated using the mechanistic AM and X-ray diffraction modeling colored by a temperature metric of interest in the diffraction volume (maximum temperature, $T_{\mathrm{Max}}$). The goal of using GPR is to learn these mappings between line profiles and underlying thermal distributions such that with a diffraction line profile, a temperature metric of interest can be extracted. Again, this places emphasis on the accuracy of the X-ray diffraction simulation, rather than the mechanistic heat transfer and fluid flow modeling. The method was recently applied to developing a mapping between diffraction line profiles and dislocation configurations within diffraction volumes \cite{Bamney2020}. In addition, the GPR approach approach shares commonalities with the Bayesian Rietveld approach introduced by Ida and Izumi \cite{ida2011application}.

GPR takes a Bayesian statistical approach to surrogate model prediction. The GPR method creates a normal distribution of mapping functions, informed by training data, with the mean of function distribution serving as a model prediction. The variance of the function distribution can serve as a confidence bound or to inform where more training data may be necessary (i.e., where there is high variance). In GPR, model output predictions (i.e., temperature metrics) are constructed from linear combinations of transformations of the input data (i.e., diffraction line profile data). The transformation and linear weights are fit according to the input training data and a chosen covariance (kernel) function. In general, the variance for a given prediction reflects the difference between the GPR model input and training data used to build the model. For example, GPR input (i.e., a diffraction line profile) that exactly matches training data will have a variance of zero, while input that is very different from the training data (e.g., a new phase is present) will produce a prediction (i.e., temperature metric) with a very high variance. As training data becomes more accurate (or at least is closer to the model input of interest), variance or uncertainty is reduced.

The most common kernel function, $k$, in GPR is the Gaussian kernel (also often referred to as the squared exponential kernel, exponentiated quadratic, or radial basis function kernel). The kernel dictates the weights used to make predictions from training data points. The Gaussian kernel takes the form
\begin{equation}
    k(\bm{x_a}, \bm{x_b}) = \sigma^2 \exp\left(-\frac{||\bm{x_a}-\bm{x_b}||^2}{2L^2}\right)
\end{equation}
where $\bm{x_a}$ and $\bm{x_b}$ are input data vectors and $\sigma^2$ is the amplitude. For this work, input data vectors are intensity values in diffraction line profiles. This kernel includes a length scale, $L$, which controls the extent of influence of a data point, affecting the variance of the function distribution. The Gaussian kernel is referenced because of its representative behavior and familiarity with the shape of Gaussian functions. As the distance between an input vector ($\bm{x_a}$) and a data point ($\bm{x_b}$) is decreased, the weight increases. Conversely, as the distance increases the weights decay in an exponential fashion.

Here, we employ the related rational quadratic kernel which is equivalent to the summation of many exponentiated quadratic kernels of different length scales:
\begin{equation}
   k(\bm{x_a}, \bm{x_b}) = \left(1+\frac{||\bm{x_a}-\bm{x_b}||^2}{2 \alpha L^2}\right)^{-\alpha}, \enspace \alpha > 0
\end{equation}
where $\alpha$ is the relative weighting between large and small length scales. Accordingly, increasing $\alpha$ reduces the amount of local variation (slows the weighting decay rate), and when $\alpha \rightarrow \infty$, the rational quadratic kernel converges to the exponentiated quadratic kernel. A range of $L$ and $\alpha$ values were tested for model training, but the closest fits to the training data (without overfitting) corresponded to $L = 1$ and $\alpha = 1$. 

\subsection{GPR Model Training and Temperature Metric Extraction}

Prior to application of the GPR models to the experimental synchrotron data, the accuracy of GPR predictions were evaluated using a set of reserved simulations. GPR models were trained using 26 of the 27 synthetic diffraction data sets, comprising 2600 images, (again nine laser parameter sets given in Table \ref{tab:test_mat} with three beam positions each) using a single processor. The reserved synthetic diffraction time series data corresponds to conditions best matching the experiment: 120 W laser power, 0.05 m/s laser speed, and placing the X-ray beam 20 $\mu$m below the top of the sample. After GPR surrogate model training and testing of the models against reserved simulated data, the trained GPR models were applied to the X-ray diffraction data collected during the synchrotron experiment. In this process, the diffraction data collected at each time step is treated as an independent data point and used as input for the various GPR models. At the end of the process, various temperature metric histories for the diffraction volume probed during the experiment are generated.




%
%

\section{Results}

\subsection{Surrogate Model Training}

\begin{figure}[h]
    \centering \includegraphics[width=1.0\textwidth]{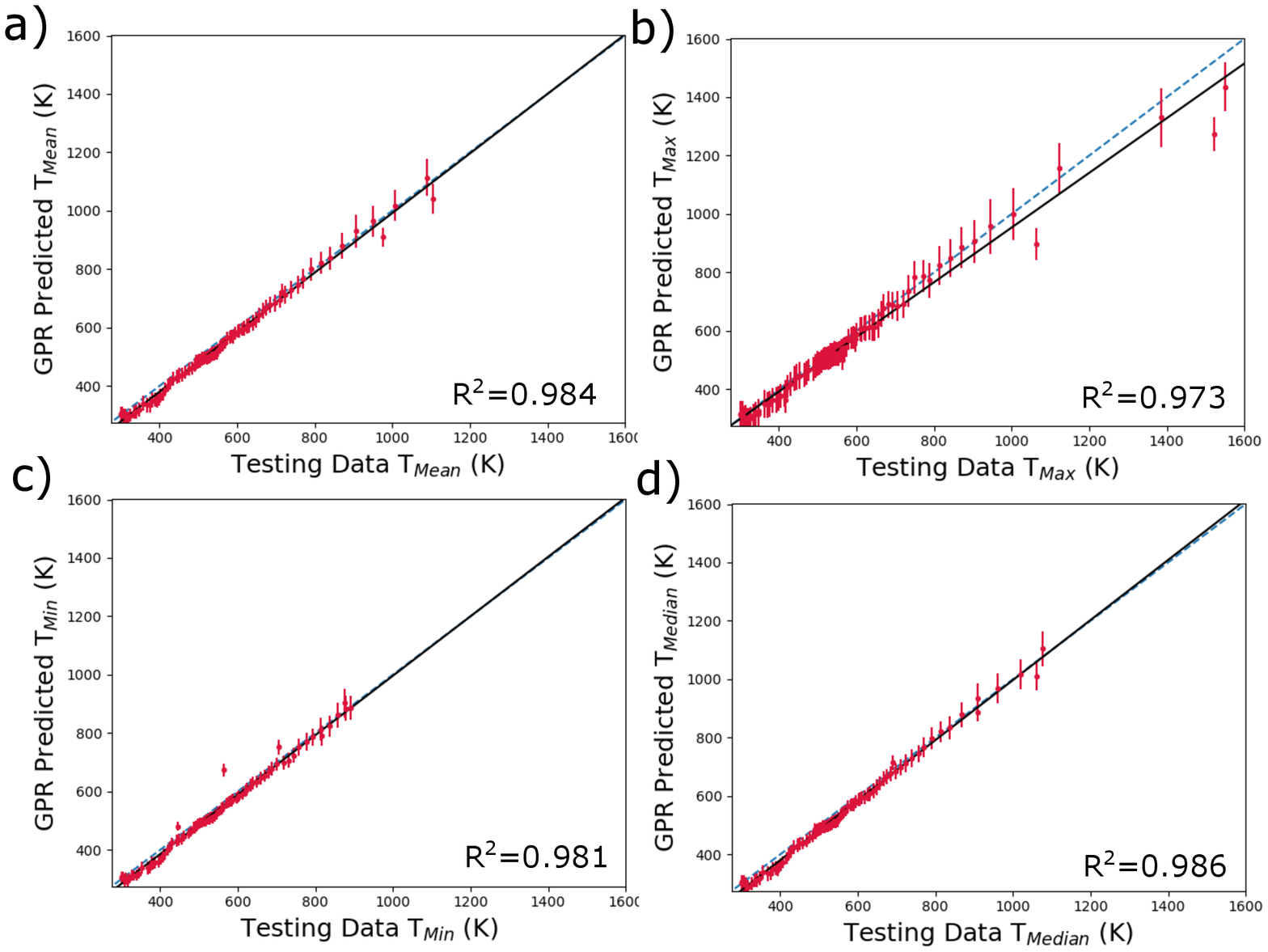}
    \caption{Comparison of prediction of temperature metrics from trained GPR surrogate models using simulated diffraction data input reserved from model training. The metrics are the a) mean $T_{\mathrm{Mean}}$, b) maximum $T_{\mathrm{Max}}$, c) minimum $T_{\mathrm{Min}}$, and d) median $T_{\mathrm{Median}}$ of the temperature distribution present in the diffraction volume. The red error bars correspond to the square root of the variance (standard deviation) of the GPR predictions. A linear regression line has been fit to the testing data and GPR predictions and is shown with a black line. The dashed blue line corresponds to perfect correlation between reserved testing data and GPR model predictions.}
    \label{fig:sim_predict}
\end{figure}

Figure \ref{fig:sim_predict} shows predictions of four temperature metrics within the diffraction volume using trained GPR models (each metric has its own trained GPR model) that use diffraction line profiles as input compared with the same temperature metrics extracted directly from the reserved thermal simulations.  The four temperature metrics are mean $T_{\mathrm{Mean}}$ (Fig. \ref{fig:sim_predict}a), maximum $T_{\mathrm{Max}}$ (Fig. \ref{fig:sim_predict}b), minimum $T_{\mathrm{Min}}$ (Fig. \ref{fig:sim_predict}c), and median $T_{\mathrm{Median}}$ (Fig. \ref{fig:sim_predict}d) temperatures. As previously described, for each GPR surrogate model prediction, the variance of the prediction associated with the normal distribution of the mapping functions can also be calculated and employed as an uncertainty. In Fig. \ref{fig:sim_predict}, the uncertainty (square root of the variance, standard deviation) of the GPR prediction is shown by the red error bars. For each temperature metric, a linear regression line was fit to the GPR predictions and is plotted with a black line. A blue dashed line corresponding to perfect correlation between the GPR model predictions and the reserved testing data is provided for comparison. The $R^2$ coefficient of determination of the GPR predictions is also provided.

As a whole, there is very good agreement between the reserved testing data and the GPR model predictions. No aphysical predictions are found across the metrics such as predictions of temperature significantly below room temperature or above the solidus temperature. It is noted that there is generally more training and testing data in the lower temperature regions due to the laser passing rapidly over the specimen, leading to generally increased uncertainty at higher temperatures (see larger red error bars). This feature is most notable for the maximum temperature predictions near the solidus temperature which has the largest uncertainties (on the order of 200 K or 15\%). Conversely, the mean temperature predictions have the smallest uncertainty (on the order of 20 K to 40 K or 6\% to 12\%) which is not surprising. The mean temperature is strongly correlated to the point of highest intensity on the diffraction peak. This can be contrasted to the minimum and maximum temperatures which will generally correspond to small volumes of material contributing to the tails of the diffraction peaks. Again, we emphasize here that these temperature metric predictions take into account contributions to peak broadening from the spatial distribution of temperature within the diffraction volume. In other words, the hottest and coldest regions may not necessarily correspond to the most extreme tail positions on the diffraction peak depending on the spatial location of the diffraction event.

\subsection{Application of GPR Surrogate Models to Experimental Data}
\label{sec:exp_data}

\begin{figure}[h]
    \centering \includegraphics[width=1.0\textwidth]{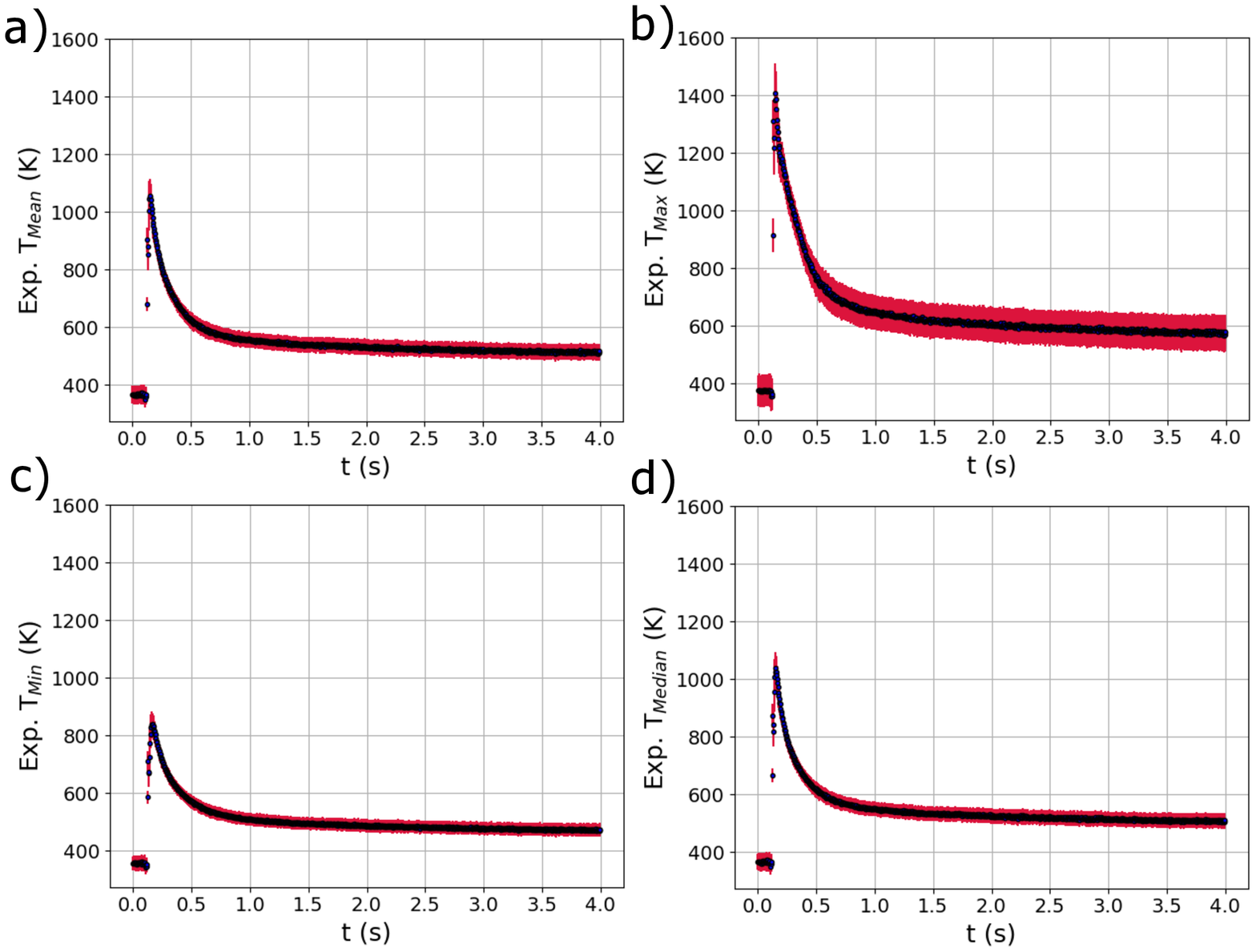}
    \caption{The evolving a) mean $T_{\mathrm{Mean}}$, b) maximum $T_{\mathrm{Max}}$, c) minimum $T_{\mathrm{Min}}$, and d) median $T_{\mathrm{Median}}$ of the distribution of temperature within the experimental X-ray diffraction volume with respect to time $t$ extracted using the trained GPR surrogate models. The red error bars correspond to the square root of the variance (standard deviation) of the GPR surrogate model predictions.}
    \label{fig:exp_predict}
\end{figure}

We have trained a series of GPR surrogate models for predicting temperature metrics from synthetic diffraction line profiles. Here we apply the trained models to analyzing experimental data captured during the synchrotron experiment described in \S \ref{sec:expdesc}. Figure \ref{fig:exp_predict} shows the evolution of mean $T_{\mathrm{Mean}}$ (Fig. \ref{fig:exp_predict}a), maximum $T_{\mathrm{Max}}$ (Fig. \ref{fig:exp_predict}b), minimum $T_{\mathrm{Min}}$ (Fig. \ref{fig:exp_predict}c), and median $T_{\mathrm{Median}}$ (Fig. \ref{fig:exp_predict}d) temperatures within the diffraction volume versus time. Again, the red error bars associated with each temperature metric measurement correspond to the uncertainty in the extracted quantity as given by the square root of variance of the GPR prediction.

The point where the moving laser passes over the diffraction volume is the clear peak in all four metrics. At its highest point, the prediction for $T_{\mathrm{Max}}$ is close to the solidus temperature for IN625 as expected, since melted material will not contribute to the diffraction peaks. Similar to the cross-validation with the simulated data, $T_{\mathrm{Max}}$ has the highest uncertainty (largest error bars). We also observe across all four metrics, that the temperature remains relatively high well after the laser has passed over the diffraction volume (on the order of 200 K above room temperature. As will be discussed, this may be due to thermomechanical stress developing within the specimen upon rapid cooling, as a positive mean stress in the volume probed will `appear' as an elevated temperature. As previously described, the uncertainty in the temperature predictions is related to how close input diffraction line profiles are to data used for GPR model training. With the transfer learning approach, if the simulations in the Source Domain used to generate the training data are missing physics, such as the development of stress due to thermal gradients, the accuracy in the Target Domain will decrease. Taking this into account, temperature metric values closer to the solidus temperature are likely the most accurate since at this point thermal expansion is largest and the mechanical stresses are lowest. This will be further discussed in \S \ref{sec:stress}. 

\section{Discussion}

Here, we have described and demonstrated a novel approach for quantifying temperature distributions within alloys during extreme heating processes that utilizes \emph{in situ} synchrotron X-ray diffraction, mechanistic modeling and X-ray simulation, and supervised machine learning. The approach was applied to quantifying temperature metrics in the bulk of an IN625 specimen during high-speed laser melting mimicking AM. The development of approaches such as that presented in this paper are important for quantifying and controlling temperature distributions during the AM build process. In turn, this information can be used to accelerate process certification and optimization of build routines to control microstructure and minimize defects. The presented method is a significant advance from current X-ray diffraction based approaches in the literature that are only capable of estimating the average temperature in an illuminated volume \cite{Hocine2020,oh2021high,oh2021microscale}. Usually analytical functions (e.g., Gaussian or Lorentzian) are fit to the diffraction line profiles and temperature is calculated from shifts of diffraction peaks. Often, to improve temperature extraction from peak fittings, experiments are conducted such that melt pools are much larger and scanning conditions are unrealistic. During this process numerous assumptions are made regarding the temperature distribution present and X-ray interaction with the sample increase the uncertainty of the estimation, most notably that all X-rays are emitted from a point source and the shape of the function used to fit the diffraction peak. In addition to decreasing accuracy with these assumptions, all information about the temperature distribution in the illuminated volume is lost.

Our effort addresses these shortcomings by directly accounting for realistic spatial thermal gradients. With these gradients accounted for, temperature metrics determined from experimental data will have increased accuracy, and we can access information about thermal gradients, a major driver of microstructure formation. Now we will examine the temperature distribution that was probed during the \emph{in situ} measurements. While the method presented is a major advance forward for bulk temperature quantification during AM, we will also discuss means to further increase accuracy and to extend the method to other quantities, such as melt-pool volume.

\subsection{Temperature Distribution Evolution}
\label{sec:dist}

\begin{figure}[h]
  \centering \includegraphics[width=0.6\textwidth]{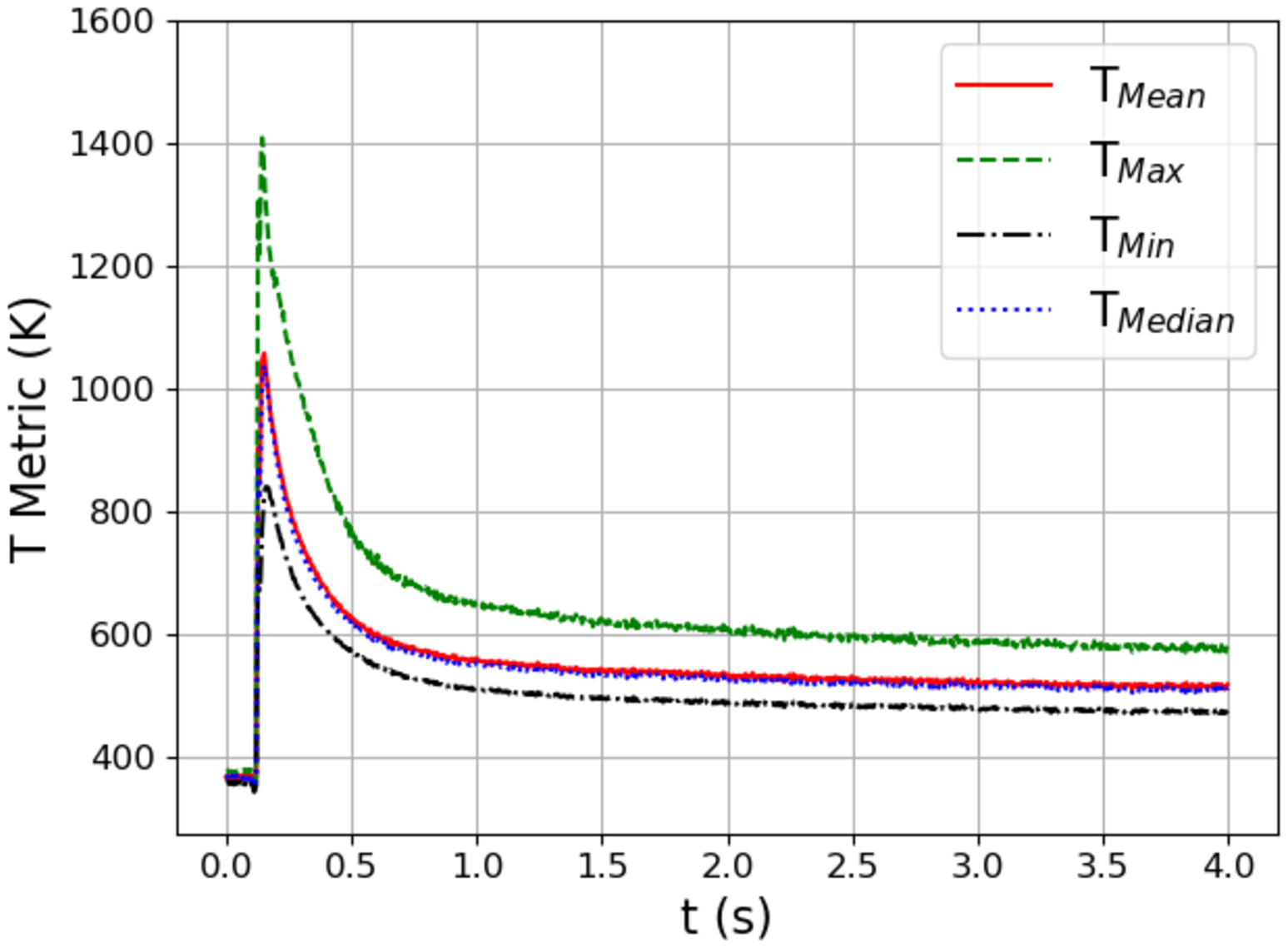}
  \caption{Comparison of the evolution of temperature $T$ metrics with time $t$ extracted from the experimental diffraction data using the GPR surrogate models.}
  \label{fig:dist}
\end{figure}

A primary benefit of the approach presented is the ability to explore the evolving \emph{distributions} of temperature present within the diffraction volume. As an example, we can analyze the distribution of temperature within the diffraction volume during the \emph{in situ} synchrotron experiment. Figure \ref{fig:dist} shows the evolution of the mean, maximum, minimum, and median temperature metrics extracted using the various GPR surrogate models together. With regards to the temperature distribution, of most interest is the range (difference of maximum and minimum) of temperatures in the crystalline phase throughout the diffraction volume. From the figure, we can see that there is nearly a 600 K difference between the maximum and minimum temperatures in the solid phase as the laser passes over the diffraction volume. We can also see the temperature difference remains throughout cooling and is still over 100 K at the end of measurement. As the mean and median temperatures are closer to the minimum temperature than the maximum, we can infer that the bulk of the diffraction volume remains cooler through thickness, matching intuition regarding localized melting. Using these data, we can also begin to establish lower bounds for the temperature gradient present. As the dimension of the diffraction volume corresponding to the specimen thickness (530 $\mu$m) is significantly larger than those defined by the incoming beam (50 $\mu$m and 30 $\mu$m), we can assume that spread of temperature is primarily along the thickness of the specimen. With this being the case, and the hottest portion of the specimen being in the center of the specimen, the lower bound of the temperature gradient can be estimated to be approximately 2250 K/mm (600K / 0.265 mm). With any liquid phase present being significantly hotter, the gradient will be larger, but a lower bound is of value for process design.

\subsection{Effects of Mechanical Elastic Strain and Stress}
\label{sec:stress}

In \S \ref{sec:exp_data}, we observed that in Fig. \ref{fig:exp_predict}, all temperature metrics appear to be converging towards predicted temperatures greater than room temperature at the end of data collection. As mentioned, this may be due to accumulation of thermomechanical stress within the specimen. Due to large thermal gradients in the material during laser melting, mechanical elastic strain and accompanying stress distributions in the material are formed to maintain deformation compatibility. These stresses can become large enough to drive plastic flow and set-up residual elastic strain and stress fields that remain in the material even upon cooling \cite{Wang2017,Phan2019,Bartlett2019}. In a cubic alloy such as IN625, tensile hydrostatic stresses and volumetric elastic strains distort the crystal lattice, and subsequently diffraction patterns, in the same fashion as increased temperature does. However, elastic strains during most mechanical deformation modes have large deviatoric components while thermal strains in cubic materials are solely volumetric. With regards to the diffraction data, heating in the absence of thermal gradients in cubic materials will cause uniform contraction or expansion of the diffraction rings. However, temperature gradients and distributions of thermal strains give rise to elastic strains and mechanical stresses to maintain deformation compatibility that, in turn, will distort diffraction rings into ellipses.

\begin{figure}[h]
  \centering \includegraphics[width=0.6\textwidth]{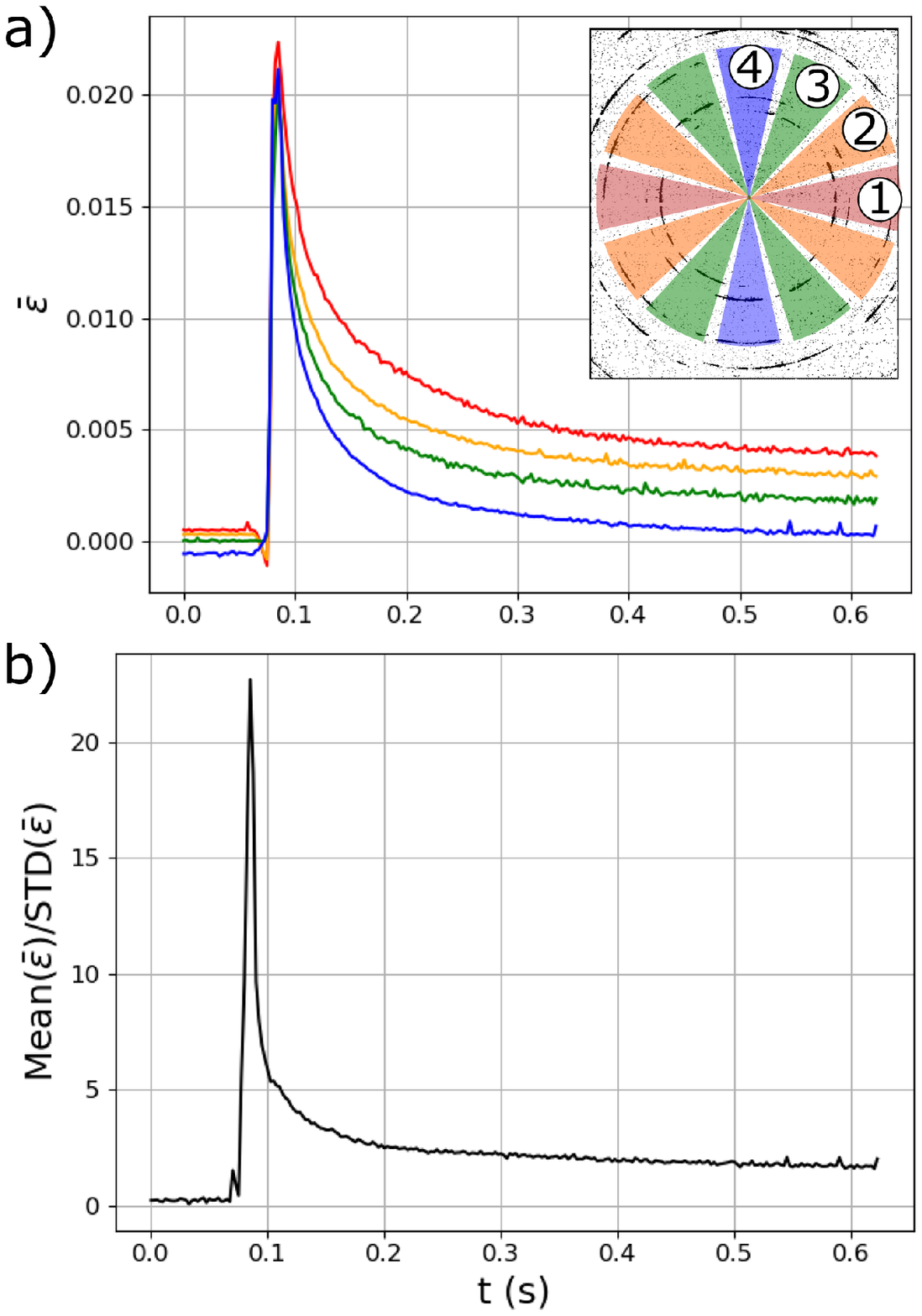}
  \caption{a) Evolution of average lattice strains $\bar{\varepsilon}$ measured from four different regions on the detector (shown in inset) through time $t$. b) Evolution of the ratio of the mean and standard deviation (STD) of the lattice strains from the four different regions.}
  \label{fig:res}
\end{figure}

To explore the role of mechanical elastic strain's effects on the temperature predictions, which are not currently included in the laser melting and diffraction simulations, the anisotropy of lattice strain around diffraction rings during the experiment were probed. Figure \ref{fig:res}a shows the evolution of average lattice strains from the first three sets of lattice planes from four different azimuthal regions around the detector. These regions are illustrated on a detector image in the inset of Fig. \ref{fig:res}a. Average lattice strains $\bar{\varepsilon}$ were first found by fitting Psuedo-Voigt peaks to the first three sets of lattice planes in each bin (noting that fits are relatively poor in the high temperature region due to peak splitting, Fig. \ref{fig:ex_comp}d). Lattice strains from each peak $\varepsilon_{hkl}$ were then determined from fit peak centers $2\theta_{hkl}$ and Bragg angles calculated from the reference lattice parameter $2\theta_{hkl0}$
\begin{equation}
    \varepsilon_{hkl}=\frac{\sin{(2\theta_{hkl0}/2)}}{(\sin{2\theta_{hkl}/2)}}-1.
\end{equation}
Average lattice strains $\bar{\varepsilon}$ from each region were calculated as an intensity weighted average of the lattice strains from each peak, calculated as
\begin{equation}
    \bar{\varepsilon}=\frac{\sum \varepsilon_{hkl} I_{hkl}}{\sum I_{hkl}}
\end{equation}
where $I_{hkl}$ is the fit maximum intensity of each peak. This averaging of lattice strains from the four different azimuthal regions assumes that the principal strain directions are nominally aligned with the sample edges and the sample coordinate system. The relatively large azimuthal regions were chosen to increase the number of grains contributing to each lattice stain measurement. In Fig. \ref{fig:res}a, we see that as the sample cools, there is a marked deviation in the lattice strains around the detector, reflecting the development of thermomechanical stresses in the specimen containing a tensile stress in the $\bm{x^S}$ direction. None of the lattice strains in any of the regions become negative, indicating still elevated temperature or a compressive stress along the beam direction $\bm{z^L}$ ($\bm{y^S}$). We also mention that there is a spread of lattice strains prior to heating that is most likely due to stresses imposed during sample mounting.

To examine an approximate measure of the ratio of volumetric to deviatoric strains caused by thermal expansion and mechanical stresses respectively, the time evolution of the ratio of the mean and standard deviation of the lattice strains from the four regions are shown in Fig. \ref{fig:res}b. In regions where the ratio is high ($>$5), the strains are dominated by volumetric (thermal) expansion and the GPR model temperature predictions are more accurate, but as the ratio gets smaller (during cooling), deviatoric (mechanical elastic) strains influence the results to produce inaccurate temperature predictions. This is reflected in the still relatively high temperature metrics determined from the GPR surrogate models in Fig. \ref{fig:dist}.

 A recent experimental effort \cite{schmeiser2021internal} has demonstrated that with a large panel area detector and capturing full diffraction rings, the effects of mechanical stresses could largely be decoupled from the effects of heating. Keeping this in mind, the method presented in this work could be extended to train GPR surrogate models with diffraction patterns including thermomechanical effects rather than just thermal effects. The GPR surrogate models would also need to be provided with independent diffraction line profiles from different azimuthal angles around the detector, but this should be readily possible. If successful, low temperature measurement accuracy will be significantly enhanced.





\subsection{Melt-Pool Volume Estimation}

\begin{figure}[h]
  \centering \includegraphics[width=0.6\textwidth]{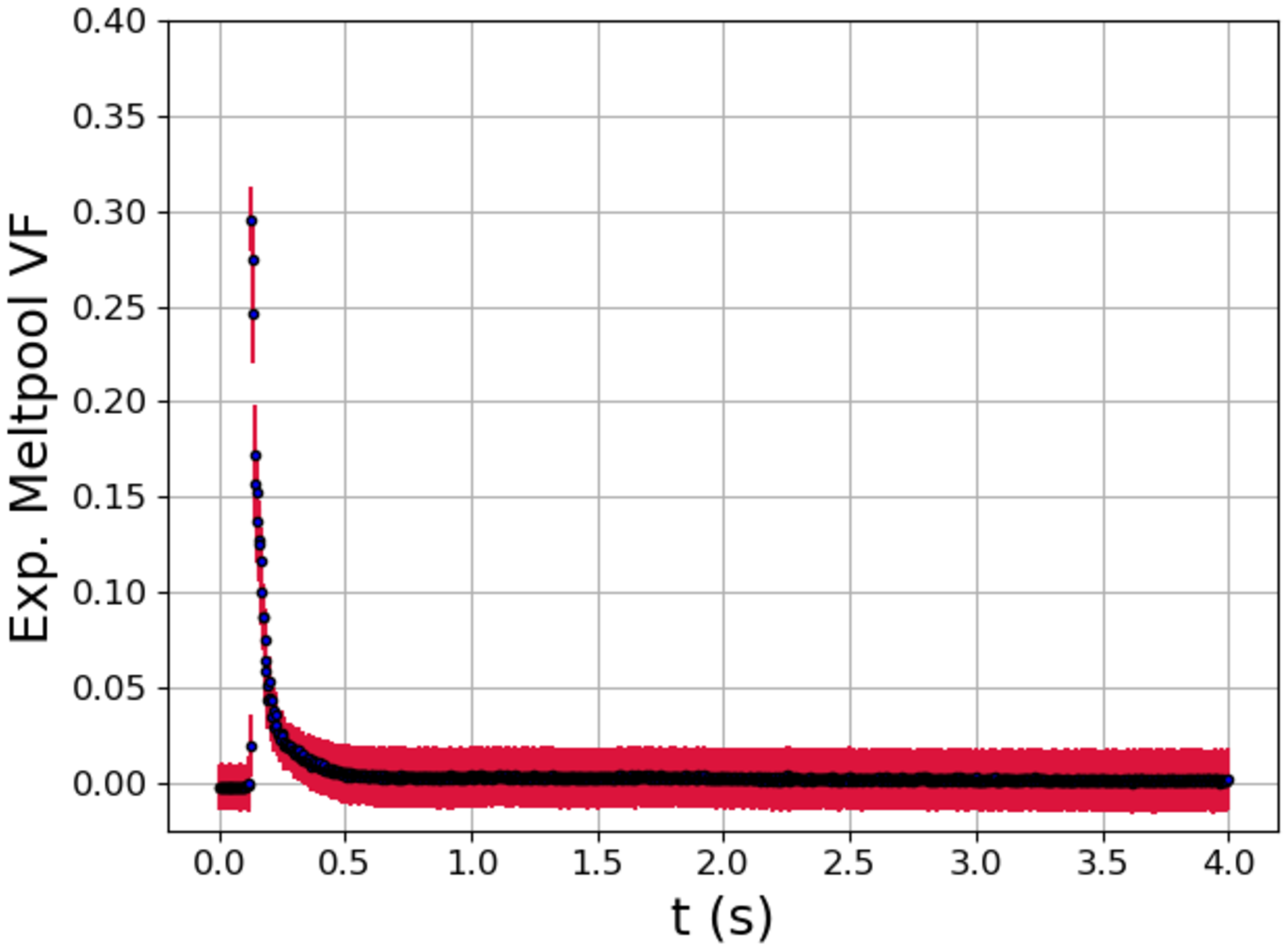}
  \caption{Estimated experimental melt-pool volume fraction (VF) versus time $t$ determined from the experimental diffraction data using a trained GPR surrogate model.}
  \label{fig:mvf}
\end{figure}

For the extraction of temperature distribution metrics, the GPR surrogate models that have been presented are primarily learning connections between diffraction peak shape and position with the temperature distributions present in the illuminated diffraction volume. As has been mentioned, when a volume of material melts, that volume will no longer contribute intensity to the measured diffraction peaks. The total integrated intensity will therefore reflect the volume of unmelted alloy and as well as the converse, the relative volume of the melt-pool. To test this idea, a final GPR surrogate model was trained in the same fashion as described above, but trained to connect diffraction line profiles to the volume fraction of the melt-pool present. This surrogate model was then provided the experimental X-ray data similarly to \S \ref{sec:exp_data}. Figure \ref{fig:mvf} shows the evolution of the melt-pool volume fraction during the experiment. We can see that the melt-pool volume fraction has a maximum of 0.3 when the laser passes over the diffraction volume. With the relatively large laser beam size (100 $\mu$m) and high power density laser conditions, a melt pool volume extending through approximately 30\% of the diffraction volume (150 $\mu$m) appears reasonable. While high-speed X-ray radiography has been used extensively in the literature to image melt-pool size, diffraction and the approach presented may provide a complimentary means to characterize sub-surface melt-pool dynamics in alloys in addition to samples with relatively little density difference between liquid and solid phases and their minimal radiographic imaging contrast.



\section{Summary and Conclusions}

For the first time, bulk maximum and minimum temperatures of the solid phase in an engineering alloy (along with temperature distribution information) are extracted from \emph{in situ} AM measurements. This was made possible by using high-fidelity mechanistic and supervised machine learning modeling to determine quantities from the experimental data, as opposed to taking a traditional approach of using experimental data to calibrate the models. The approach consists of training Gaussian Process Regression surrogate models using a combination of heat transfer and fluid flow modeling and X-ray diffraction modeling. Each surrogate model links diffraction line profiles to metrics describing the temperature distribution present within the diffraction volumes including maximum, minimum, mean, and median temperature. The smallest uncertainties determined from the GPR models are approximately 5\% for the minimum, median, and mean temperatures. In contrast, the largest for the maximum temperature (approximately 15\%), near the solidus temperature. The trained surrogate models were successfully applied to extracting these metrics from \emph{in situ} high-energy synchrotron diffraction data collected during additive manufacturing (laser melting) of a thin wall of Inconel 625. Despite the larger uncertainties from the GPR surrogate model output, temperature metric accuracy is believed to be greatest in the high temperature regime and decreases in the low temperature regime when applied to experimental data, as stress develops and distorts the material present. The discussion described future efforts to account for thermomechanical stress formation to increase model accuracy and extend the approach to extract other information about the volume probed including the volume of the melt-pool in the diffraction volume.

\section*{Acknowledgements}
	REL and DCP were supported by the National Institute of Standards and Technology under award 70NANB20H208. The work of TQP was performed under the auspices of the U.S. Department of Energy by Lawrence Livermore National Laboratory under Contract DE-AC52-07NA27344. Computations for this research were performed on the Pennsylvania State University's Institute for Computational and Data Sciences' Roar supercomputer. This research used resources of the Advanced Photon Source, a U.S. Department of Energy (DOE) Office of Science user facility operated for the DOE Office of Science by Argonne National Laboratory under Contract No. DE-AC02-06CH11357. Mr. Ian Wietecha-Reiman is thanked for help performing the dilatometry measurements and Mr. Kenneth Peterson is thanked for performing the electron backscatter diffraction measurements.


\section*{Data and Code Availability Statement}

All data used for this work is available upon reasonable request. The Python-based diffraction simulation and GPR training codes are also available upon request.

\newpage
\clearpage

\bibliographystyle{elsarticle-num}
\bibliography{References.bib}

\begin{thebibliography}{10}
\expandafter\ifx\csname url\endcsname\relax
  \def\url#1{\texttt{#1}}\fi
\expandafter\ifx\csname urlprefix\endcsname\relax\def\urlprefix{URL }\fi
\expandafter\ifx\csname href\endcsname\relax
  \def\href#1#2{#2} \def\path#1{#1}\fi

\bibitem{moylan2014infrared}
S.~Moylan, E.~Whitenton, B.~Lane, J.~Slotwinski, Infrared thermography for
  laser-based powder bed fusion additive manufacturing processes, in: AIP
  Conference Proceedings, Vol. 1581, American Institute of Physics, 2014, pp.
  1191--1196.

\bibitem{fox2017measurement}
J.~C. Fox, B.~M. Lane, H.~Yeung, Measurement of process dynamics through
  coaxially aligned high speed near-infrared imaging in laser powder bed fusion
  additive manufacturing, in: Thermosense: thermal infrared applications XXXIX,
  Vol. 10214, SPIE, 2017, pp. 34--50.

\bibitem{fisher2018toward}
B.~A. Fisher, B.~Lane, H.~Yeung, J.~Beuth, Toward determining melt pool quality
  metrics via coaxial monitoring in laser powder bed fusion, Manufacturing
  letters 15 (2018) 119--121.

\bibitem{montazeri2019heterogeneous}
M.~Montazeri, A.~R. Nassar, C.~B. Stutzman, P.~Rao, Heterogeneous sensor-based
  condition monitoring in directed energy deposition, Additive Manufacturing 30
  (2019) 100916.

\bibitem{dunbar2018assessment}
A.~J. Dunbar, A.~R. Nassar, Assessment of optical emission analysis for
  in-process monitoring of powder bed fusion additive manufacturing, Virtual
  and Physical Prototyping 13~(1) (2018) 14--19.

\bibitem{forien2020detecting}
J.-B. Forien, N.~P. Calta, P.~J. DePond, G.~M. Guss, T.~T. Roehling, M.~J.
  Matthews, Detecting keyhole pore defects and monitoring process signatures
  during laser powder bed fusion: A correlation between in situ pyrometry and
  ex situ x-ray radiography, Additive Manufacturing 35 (2020) 101336.

\bibitem{ashby2022thermal}
A.~Ashby, G.~Guss, R.~K. Ganeriwala, A.~A. Martin, P.~J. DePond, D.~J. Deane,
  M.~J. Matthews, C.~L. Druzgalski, Thermal history and high-speed optical
  imaging of overhang structures during laser powder bed fusion: A
  computational and experimental analysis, Additive Manufacturing 53 (2022)
  102669.

\bibitem{Kenel2016}
C.~Kenel, D.~Grolimund, J.~L. Fife, V.~A. Samson, S.~{Van Petegem}, H.~{Van
  Swygenhoven}, C.~Leinenbach, {Combined in situ synchrotron micro X-ray
  diffraction and high-speed imaging on rapidly heated and solidified Ti-48Al
  under additive manufacturing conditions}, Scripta Materialia 114 (2016)
  117--120.
\newblock \href {https://doi.org/10.1016/j.scriptamat.2015.12.009}
  {\path{doi:10.1016/j.scriptamat.2015.12.009}}.

\bibitem{Calta2018}
N.~P. Calta, J.~Wang, A.~M. Kiss, A.~A. Martin, P.~J. Depond, G.~M. Guss,
  V.~Thampy, A.~Y. Fong, J.~N. Weker, K.~H. Stone, et~al., An instrument for in
  situ time-resolved x-ray imaging and diffraction of laser powder bed fusion
  additive manufacturing processes, Review of Scientific Instruments 89~(5)
  (2018) 055101.

\bibitem{Cunningham2019}
R.~Cunningham, C.~Zhao, N.~Parab, C.~Kantzos, J.~Pauza, K.~Fezzaa, T.~Sun,
  A.~D. Rollett, Keyhole threshold and morphology in laser melting revealed by
  ultrahigh-speed x-ray imaging, Science 363~(6429) (2019) 849--852.

\bibitem{Hocine2020}
S.~Hocine, H.~{Van Swygenhoven}, S.~{Van Petegem}, C.~S.~T. Chang,
  T.~Maimaitiyili, G.~Tinti, D.~{Ferreira Sanchez}, D.~Grolimund, N.~Casati,
  {Operando X-ray diffraction during laser 3D printing}, Materials Today 34
  (2020) 30--40.
\newblock \href {https://doi.org/10.1016/j.mattod.2019.10.001}
  {\path{doi:10.1016/j.mattod.2019.10.001}}.

\bibitem{oh2021high}
S.~A. Oh, R.~E. Lim, J.~W. Aroh, A.~C. Chuang, B.~J. Gould, B.~Amin-Ahmadi,
  J.~V. Bernier, T.~Sun, P.~C. Pistorius, R.~M. Suter, et~al., High speed
  synchrotron x-ray diffraction experiments resolve microstructure and phase
  transformation in laser processed ti-6al-4v, Materials Research Letters
  9~(10) (2021) 429--436.

\bibitem{oh2021microscale}
S.~A. Oh, R.~E. Lim, J.~W. Aroh, A.~C. Chuang, B.~J. Gould, J.~V. Bernier,
  N.~Parab, T.~Sun, R.~M. Suter, A.~D. Rollett, Microscale observation via
  high-speed x-ray diffraction of alloy 718 during in situ laser melting, JOM
  73~(1) (2021) 212--222.

\bibitem{thampy2020subsurface}
V.~Thampy, A.~Y. Fong, N.~P. Calta, J.~Wang, A.~A. Martin, P.~J. Depond, A.~M.
  Kiss, G.~Guss, Q.~Xing, R.~T. Ott, et~al., Subsurface cooling rates and
  microstructural response during laser based metal additive manufacturing,
  Scientific reports 10~(1) (2020) 1--9.

\bibitem{Bamney2020}
D.~Bamney, A.~Tallman, L.~Capolungo, D.~E. Spearot, Virtual diffraction
  analysis of dislocations and dislocation networks in discrete dislocation
  dynamics simulations, Computational Materials Science 174 (2020) 109473.

\bibitem{son2020creep}
K.-T. Son, M.~E. Kassner, K.~A. Lee, The creep behavior of additively
  manufactured inconel 625, Advanced Engineering Materials 22~(1) (2020)
  1900543.

\bibitem{levine2020outcomes}
L.~Levine, B.~Lane, J.~Heigel, K.~Migler, M.~Stoudt, T.~Phan, R.~Ricker,
  M.~Strantza, M.~Hill, F.~Zhang, et~al., Outcomes and conclusions from the
  2018 am-bench measurements, challenge problems, modeling submissions, and
  conference, Integrating Materials and Manufacturing Innovation 9 (2020)
  1--15.

\bibitem{Zhao2017}
C.~Zhao, K.~Fezzaa, R.~W. Cunningham, H.~Wen, F.~De~Carlo, L.~Chen, A.~D.
  Rollett, T.~Sun, Real-time monitoring of laser powder bed fusion process
  using high-speed x-ray imaging and diffraction, Scientific reports 7~(1)
  (2017) 1--11.

\bibitem{weiss2016survey}
K.~Weiss, T.~M. Khoshgoftaar, D.~Wang, A survey of transfer learning, Journal
  of Big data 3~(1) (2016) 1--40.

\bibitem{Mukherjee2018}
T.~Mukherjee, H.~L. Wei, A.~De, T.~DebRoy, {Heat and fluid flow in additive
  manufacturing—Part I: Modeling of powder bed fusion}, Computational
  Materials Science 150~(April) (2018) 304--313.
\newblock \href {https://doi.org/10.1016/j.commatsci.2018.04.022}
  {\path{doi:10.1016/j.commatsci.2018.04.022}}.

\bibitem{Mukherjee2018a}
T.~Mukherjee, H.~L. Wei, A.~De, T.~DebRoy, {Heat and fluid flow in additive
  manufacturing – Part II: Powder bed fusion of stainless steel, and
  titanium, nickel and aluminum base alloys}, Computational Materials Science
  150~(April) (2018) 369--380.
\newblock \href {https://doi.org/10.1016/j.commatsci.2018.04.027}
  {\path{doi:10.1016/j.commatsci.2018.04.027}}.

\bibitem{Pagan2020}
D.~C. Pagan, K.~K. Jones, J.~V. Bernier, T.~Q. Phan, {A Finite Energy
  Bandwidth-Based Diffraction Simulation Framework for Thermal Processing
  Applications}, Jom 72~(12) (2020) 4539--4550.
\newblock \href {https://doi.org/10.1007/s11837-020-04443-7}
  {\path{doi:10.1007/s11837-020-04443-7}}.

\bibitem{Bernier2011}
J.~V. Bernier, N.~R. Barton, U.~Lienert, M.~P. Miller, {Far-field high-energy
  diffraction microscopy: a tool for intergranular orientation and strain
  analysis}, Journal of Strain Analysis for Engineering Design 46~(7) (2011)
  527--547.
\newblock \href {https://doi.org/10.1177/0309324711405761}
  {\path{doi:10.1177/0309324711405761}}.

\bibitem{pawel1985survey}
R.~Pawel, R.~Williams, Survey of physical property data for several alloys,
  Tech. rep., Oak Ridge National Lab. (1985).

\bibitem{Williams2006}
C.~K. Williams, C.~E. Rasmussen, Gaussian processes for machine learning,
  Vol.~2, MIT press Cambridge, MA, 2006.

\bibitem{pedregosa2011scikit}
F.~Pedregosa, G.~Varoquaux, A.~Gramfort, V.~Michel, B.~Thirion, O.~Grisel,
  M.~Blondel, P.~Prettenhofer, R.~Weiss, V.~Dubourg, et~al., Scikit-learn:
  Machine learning in python, Journal of machine learning research 12~(Oct)
  (2011) 2825--2830.

\bibitem{ida2011application}
T.~Ida, F.~Izumi, Application of a theory for particle statistics to structure
  refinement from powder diffraction data, Journal of Applied Crystallography
  44~(5) (2011) 921--927.

\bibitem{Wang2017}
Z.~Wang, E.~Denlinger, P.~Michaleris, A.~D. Stoica, D.~Ma, A.~M. Beese,
  {Residual stress mapping in Inconel 625 fabricated through additive
  manufacturing: Method for neutron diffraction measurements to validate
  thermomechanical model predictions}, Materials and Design 113 (2017)
  169--177.
\newblock \href {https://doi.org/10.1016/j.matdes.2016.10.003}
  {\path{doi:10.1016/j.matdes.2016.10.003}}.

\bibitem{Phan2019}
T.~Q. Phan, M.~Strantza, M.~R. Hill, T.~H. Gnaupel-Herold, J.~Heigel, C.~R.
  D’Elia, A.~T. DeWald, B.~Clausen, D.~C. Pagan, J.~P. Ko, et~al., Elastic
  residual strain and stress measurements and corresponding part deflections of
  3d additive manufacturing builds of in625 am-bench artifacts using neutron
  diffraction, synchrotron x-ray diffraction, and contour method, Integrating
  Materials and Manufacturing Innovation 8~(3) (2019) 318--334.

\bibitem{Bartlett2019}
J.~L. Bartlett, X.~Li, {An overview of residual stresses in metal powder bed
  fusion}, Additive Manufacturing 27~(January) (2019) 131--149.
\newblock \href {https://doi.org/10.1016/j.addma.2019.02.020}
  {\path{doi:10.1016/j.addma.2019.02.020}}.

\bibitem{schmeiser2021internal}
F.~Schmeiser, E.~Krohmer, N.~Schell, E.~Uhlmann, W.~Reimers, Internal stress
  evolution and subsurface phase transformation in titanium parts manufactured
  by laser powder bed fusion—an in situ x-ray diffraction study, Advanced
  Engineering Materials 23~(11) (2021) 2001502.

\end{thebibliography}

\end{document}